\documentclass[useAMS,usenatbib]{mn2e}

\usepackage[dvips]{graphicx}

\newcommand{\reduceme}{\mbox{R\raisebox{-0.35ex}{E}D%
\hspace{-0.05em}\raisebox{0.85ex}{uc}\hspace{-0.90em}%
\raisebox{-.35ex}{{m}}\hspace{0.05em}E}}

\title[Stellar populations of massive elliptical galaxies in very rich clusters]{Stellar populations of massive elliptical galaxies in very rich clusters}
\author[C. Carretero, A. Vazdekis and J. E. Beckman]{C. Carretero$^{1}$\thanks{E-mail: cch@iac.es}, A. Vazdekis$^{1}$ and J. E. Beckman$^{1, 2}$\\
$^{1}$Instituto de Astrof\'{\i}sica de Canarias, V\'{\i}a L\'actea s/n, 38200 La Laguna, Tenerife, Spain\\
$^{2}$Consejo Superior de Investigaciones Cient\'{\i}ficas, 28006 Madrid, Spain}
\begin{document}

\date{Accepted 2006 December 1. Received 2006 November 30; in original form 2006 September 29}

\pagerange{\pageref{firstpage}--\pageref{lastpage}} \pubyear{2006}

\maketitle

\label{firstpage}

\begin{abstract}

We present a detailed stellar population analysis of 27 massive elliptical galaxies within 4 very rich clusters at redshift $z\sim0.2$: A115, A655, A963 and A2111. Using the new, high-resolution stellar populations models developed in our group, we obtained accurate estimates of the mean luminosity-weighted ages and relative abundances of CN, Mg and Fe. We derived the age, [CN/H], [Mg/H], [Fe/H] and the abundance ratios [CN/Fe] and [Mg/Fe] as functions of the galaxy velocity dispersion, $\sigma$. We have found that [CN/H] and [Mg/H] are correlated with $\sigma$ while [Fe/H] and Log(age) are not. In addition, both abundance ratios [CN/Fe] and [Mg/Fe] increase with $\sigma$. Furthermore, the [CN/H]$-\sigma$ and [CN/Fe]$-\sigma$ slopes are steeper for galaxies in very rich clusters than those in the less dense Virgo and Coma clusters. On the other hand, [Mg/H]$-\sigma$ and [Mg/Fe]$-\sigma$ slopes keep constant as functions of the environment. Our results are compatible with a scenario in which the stellar populations of massive elliptical galaxies, independently of their environment and mass, had formation timescales shorter than $\sim1$~Gyr. This result implies that massive elliptical galaxies have evolved passively since, at least, as long ago as $z\sim2$. For a given galaxy mass the duration of star formation is shorter in those galaxies belonging to more dense environments; whereas the mass-metallicity relation appears to be also a function of the cluster properties: the denser the environment is, the steeper are the correlations. Finally, we show that the abundance ratios [CN/Fe] and [Mg/Fe] are the key ``chemical clocks'' to infer the star formation history timescales in ellipticals. In particular, [Mg/Fe] provides an upper limit for those formation timescales, while [CN/Fe] apperars to be the most suitable parameter to resolve them in elliptical galaxies with $\sigma<300$~km~s$^{-1}$.

\end{abstract}


\section{Introduction}

Stellar Population analysis has recently shown increasing power in constraining galaxy formation scenarios. This follows the relative success of CDM cosmologies, giving rise to the hierarchical build-up of galaxies as a consequence of the initial conditions \citep[e.g.][]{press74,white91,somerville99,delucia06}. Many structural and dynamical properties of galaxies in clusters are explained in this scenario, though problems remain: for instance, the absence of the predicted mass cusps in the centres of ellipticals and bulges \citep[e.g.][]{valenzuela05}, and the prediction of far more satellite galaxies than those observed \citep{klypin99,moore99}.

Stellar populations offer a fossil record of the star formation and chemical evolution of galaxies, most clearly in elliptical galaxies, thought to have formed by the merging of discs, so stellar population studies provide very strong constraints on these galaxy formation scenarios. For example, it is particularly hard to reconcile the hierarchical models with the result that massive galaxies show significantly older mean luminosity weighted ages than their smaller counterparts \citep{kauffmann03,yamada06}. 

A basis for population analysis is the colour-magnitude relation (CMR). \citet{sandage72} showed that luminous elliptical galaxies are redder than fainter ones. The monolithic explanation is that more massive galaxies produce more metals at an early phase \citep{kodama97,chiosi02,matteucci03}, implying old stellar populations in elliptical galaxies. On the contrary, in the hierarchical explanation the metals are produced over an extended period in merger induced star formation, so an intermediate age population should be detectable \citep{delucia06}. The results so far are ambiguous: \citet{terlevich99} found evidence for the former, while \citet{jorgensen99} and \citet{trageretal00} for the latter.

It is hard to separate age from metallicity effects using the CMR \citep{wortheyetal94} and this bedevils all analyses of unresolved stellar populations. Many studies of clusters have used the pioneering Lick/IDS spectral indices \citep{wortheyetal94} whose reliablity is limited by their resolution dependence as uncertain corrections for broadening and instrumental effects are needed \citep{worthey97}. One can make some advance by using modelled integrated spectra, which can be broadened to match the galaxy velocity dispersion, sigma (although indices are still used in this method, one has more flexibility that with the fitting functions required in the earlier technique). Models employing the use of integrated spectra at relatively high resolution  have been developed by \citet{vazdekis99} which allow one to effect a better separation of age and metallicity than with the Lick indices \citep{vazdekisarimoto99,vazdekisetal01b,kuntschneretal02}. Another advantage we have in the present work is that we have been able to employ the new stellar population synthesis models of Vazdekis  et al. (2006, in preparation) which are improved versions of the earlier \citet{vazdekis99} models, based on a greatly extended stellar spectral library , MILES \citep{sanchezblazquez06miles}. The new library contains spectra of over 1000 stars, convering unequalled ranges in stellar metallicity and spectral class in the wavelength range lambdalambda $\lambda\lambda~3500-7500$ \AA, at a resolution of 2.3~\AA~(FWHM).

This new generation of models has shown considerable ability to break the age metallicity degeneracy and has been succesfully applied to a number of elliptical galaxies selected using the CMR of the Virgo cluster. Once the age metallicity degeneracy is resolved, it is fairly easy to derive the abundances of individual elements using these models. For example, \citet{vazdekisetal01b} have found that luminous and moderately-luminous ellipticals ($M_V$$<$$-18.3$~mag; $\sigma \ge 100$~km~s$^{-1}$) are essentially old (t $>$ 8-10 Gyr) without a clear trend with luminosity. [Mg/H] correlates tightly with velocity dispersion, while [Fe/H] is showing only a modest trend, implying that the CMR is driven essentially by metallicity (in particular, by the abundance of the $\alpha$-enhanced elements) and that star formation in massive ellipticals virtually had stopped before  SNIa started to contaminate the ISM with iron-group elements. In short, star formation was intense and short-lived ($\le1$ Gyr). Admittedly, however, the number of ellipticals observed in this way is small and the results may not be statistically significant. Contrary to these findings, \citet{trager05} found a few young moderately massive ellipticals in Coma. \citet{caldwell03} analyzed  integrated spectra of 175 nearby early-type galaxies (mixture of field and Virgo E, dE, S0, and dS0's) to find a trend between the population age and the velocity dispersion, such that low-$\sigma$ galaxies have younger luminosity-weighted mean ages. It is therefore not yet clear if cluster ellipticals with $\sigma \ge 100$~km~s$^{-1}$ are uniformly old or are progressively younger towards less massive objects \citep{bernardi06}.

One of the key pieces of information is to see how abundance ratios in early-type galaxies vary with the cluster masses (as sampled via their richness class or their X-ray luminosities, for example) and, in this way, obtain vital insights into galaxy assembly timescales, as functions of their environment \citep[e.g.][]{worthey98,carretero04}. In particular, overabundances of [Mg/Fe] compared with the solar ratio have been found in massive elliptical galaxies \citep{peletier89,worthey92,vazdekisetal97}. These have been interpreted via several possible scenarios based on the fact that Mg is mainly produced in Type II supernovae \citep{faberetal92,matteucci94}, and include different star formation rates (SFR) and a time dependent initial mass function (IMF).

Based on SDSS data, we followed this approach previously \citep{carretero04}, testing the evolution of the ratios between CN, Mg and Fe as a function of the cluster X-ray luminosity. We tentatively concluded that early-type galaxies are fully assembled on timescales around the epoch of the massive release of CN into the ISM. This timescale is shorter than predicted by classical hierarchical models \citep[e.g.][]{kauffmanncharlot98}, and thus it is vital to test this conclusion.

In this paper, we concentrate on the stellar population analysis of 27 elliptical galaxies distributed in 4 rich, intermediate-redshift ($z\sim0.2$) Abell clusters: A115, A655, A963 and A2111. Selected cluster galaxies have been morphologically and photometrically studied in the literature, so we focused on those galaxies already classified as ellipticals. We determined mean luminosity-weighted ages and a number of abundance ratios of species such as Fe, CN and Mg; and their relation with the velocity dispersion of the galaxies. Finally, we compared our results with those obtained by other authors in galaxies belonging to less rich clusters such as Virgo and Coma.

The outline of the paper is as follows: Section 2 presents the observations, sample selection and data reduction. Section 3 shows the relations found between the abundances of selected elements and the velocity dispersion, and the comparison with results obtained in less dense environments. Section 4 presents the discussion of our results. Finally, Section 5 summarizes our conclusions.

Where necessary, we adopt a cosmological model with H$_0$=71 km s$^{-1}$ Mpc$^{-1}$, $\Omega_m$=0.3 and $\Omega_\Lambda$=0.7.

\begin{figure}
\resizebox{\hsize}{!}{\includegraphics{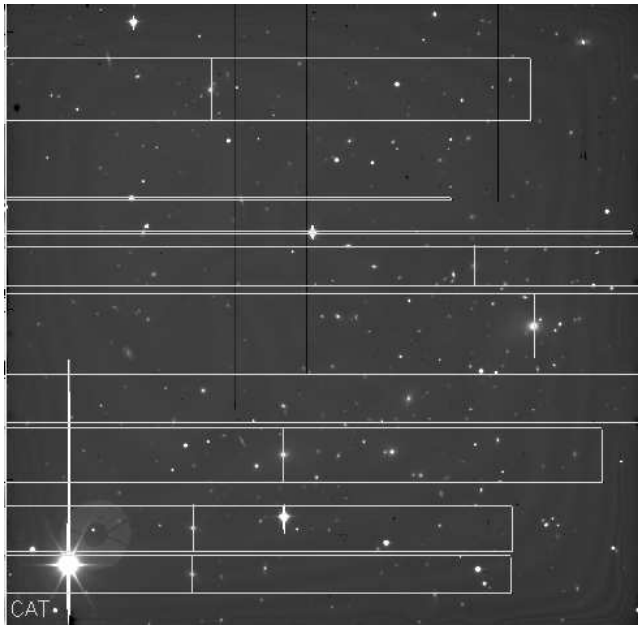}}
\caption{Field of Abell 655. We have superimposed the MOS mask \#1 used during our observations of this cluster.}
\label{mask}
\end{figure}

\section{Observations and Data}

\subsection{Sample selection}

We have studied a set of 27 elliptical galaxies belonging to 4 very rich \citep[richness class 3 and $L_{\rm X}\sim10^{45}$~erg~s$^{-1}$;][]{ebelingetal98} Abell clusters: A115, A655, A963 and A2111. Table \ref{clusterproperties} presents detailed information on each cluster. We have selected these clusters because they represent very dense environments, since our aim is to compare the results of the stellar population analysis performed on galaxies in very rich clusters with the results already obtained in less rich clusters, such as Virgo \citep[richness class 1 and $L_{\rm X}=0.3\times10^{44}$~erg~s$^{-1}$;][]{ledlowetal03} or Coma \citep[richness class 2 and $L_{\rm X}=1.8\times10^{44}$~erg~s$^{-1}$;][]{ledlowetal03} clusters.

A115 presents 2 main X-ray substructures, which suggests that this cluster is not relaxed but in an intermediate dynamical stage of merging prior to virialization \citep{beers83,gutierrez05}. A similar scenario can be traced for A655, which presents various substructures \citep{strazzullo05}, and for A2111, which shows a double-peaked core \citep{jeltema05}. A very different situation is found for A963 which, even though a high $L_{\rm X}$ cluster, is a quite relaxed system with some possible very small-scale structure in the core \citep{jeltema05}.

\begin{table*}
\begin{minipage}{150mm}
\caption{Logbook of the observations and main properties of selected Abell clusters. X-ray luminosity was obtained from \citet{ebelingetal98}. N$_{\rm gal}$ column contains the number of galaxies studied for each cluster. `` Ref.'' column indicates the reference from which we have obtained the morphology of analysed galaxies: 1: \citet{rakos00}; 2: \citet{flin95}; 3: \citet{balogh02}; 4: \citet{fasano00}.}
\label{clusterproperties}
\begin{tabular}{cccccccccc}
\hline
Cluster & Date & $t_{\rm exp}$ & $\alpha$ & $\delta$ & $z$ & Richness & $L_{\rm X}$ & N$_{\rm gal}$ & Ref. \\
        &      &     (s)         &  (J2000.0) & (J2000.0)  &     &   class  & $(10^{44}$erg s$^{-1})$ & &  \\
\hline
A115  & Oct 2005   & 16200                &  00 55 59.5  & +26 19 14  & 0.1971 & 3 & 14.59 &  6 & 1 \\
A655  & Jan 2005   & 12600\footnote{$t_{\rm exp}$ per mask (2 in total).}  &  08 25 20.2  & +47 07 13  & 0.1265 & 3 &  6.73 & 10 & 2 \\
A963  & Jan 2005   & 16200	          &  10 17 09.6  & +39 01 00  & 0.2060 & 3 & 10.41 &  6 & 3 \\
A2111 & May 2005\footnote{The observing program was interrupted for several hours during two nights due to the occurrence of 2 GRB warnings: GRB 050505 and GRB 050508.}   & 18000   &  15 39 38.3  & +34 24 21  & 0.2290 & 3 & 10.94 &  5 & 4 \\
\hline
\end{tabular}
\end{minipage}
\end{table*}

The selected galaxies are the brightest galaxies in their respective clusters ($15.1 < m_V < 17.2$) and have been already morphologically analyzed in the literature \citep[see Table \ref{clusterproperties} for details;][]{flin95,fasano00,rakos00,balogh02}. Thanks to this previous work we were able to focus our observations on those galaxies already classified as ellipticals. The velocity dispersion values of the galaxies in our sample cover the range 220~km~s$^{-1}<\sigma<450$~km~s$^{-1}$.

\subsection{Observations}

Multi-slit spectroscopy was carried out with the DOLORES spectrograph on the 3.5m Telescopio Nazionale Galileo at the Roque de los Muchachos Observatory (La Palma, Spain) in 3 runs. Table \ref{clusterproperties} details each observing run. DOLORES CCD scale is 0.275 arcsec~pix$^{-1}$ which yields a field of view of about $9.4\times9.4$ arcmin. We designed Multi Object Spectroscopy (MOS) masks containing $\sim6$ slits each. As an example, Figure \ref{mask} shows the field of A655. We have overimposed the image of one of the 2 MOS masks used to obtain the spectra from the galaxies of this cluster.

For all masks, in all observing runs, we used the same instrumental setup: the medium-resolution MR-B\#2 grism, in the wavelength range $\lambda\lambda$~s3500-7000~\AA, and 1.1$''$ width slits. The dispersion was 1.7~\AA~pix$^{-1}$ with an instrumental resolution of 8 \AA\ (FWHM).

We performed exposures of 1800 seconds, interspersed with He+Ar arc-lamp spectra in order to correct for possible changes in the wavelength calibration during the night. Table \ref{clusterproperties} presents the total exposure time for each mask. In addition, we observed a number of G- and K-type stars to be used as templates for velocity dispersion measurements. Finally, flux standards from \citet{oke90} were observed to correct the continuum shape of the spectra.

\subsection{Data Reduction}

The standard data reduction procedures (flat-fielding, wavelength calibration, sky subtraction and fluxing) were performed with IRAF packages. In addition, we used \reduceme\ \citep{cardiel99} to remove cosmic rays, since this reduction package allows full control of this process.

Once reduced, we coadded our 2D spectra in the spatial direction to obtain an average value of S/N$\sim40$ in the full sample. Considering the physical distance to each cluster, we integrated the light of the galaxies out to a galactocentric radius comparable to the galaxies already analyzed in Virgo and Coma clusters \citep{vazdekistrujillo04,sanchezblazquezetal06}. In particular, we considered an equivalent aperture of $5''$ at a redshift $z=0.016$. This way, we were able to establish direct comparisons between the our spectral index measurements and those performed in these clusters.

\begin{figure}
\resizebox{\hsize}{!}{\includegraphics{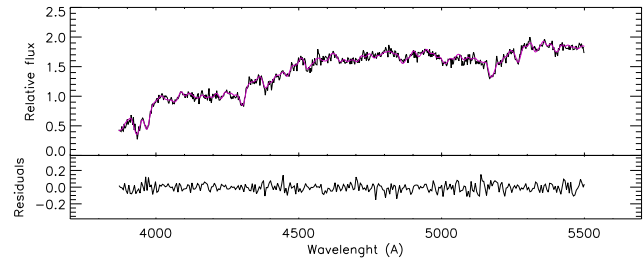}}
\vspace{0.5cm}
\caption{{\em Top}: An example of one of our spectra. The black line corresponds to the observed spectrum of a representative galaxy of A115 ($\sigma=330$~km~s$^{-1}$). The red line corresponds to a model of Vazdekis et al. (2006, in preparation) with $t=15$~Gyr and solar metallicity. {\em Bottom}: Residual after subtracting the model from the observed spectrum.}
\label{espectros}
\end{figure}

Figure \ref{espectros} shows an example of the final spectrum of a galaxy in A115 with $\sigma=330$~km~s$^{-1}$. We have overplotted a synthetic spectrum of a model of age 15~Gyr and solar metallicity. The bottom panel presents the residuals obtained when subtracting the model spectrum from the observed one.

\section{Analysis and Results}

Once the reduced spectra for all the galaxies in our sample were obtained, we proceeded to measure their velocity dispersion and different spectral indices, which are the basis of our stellar population analysis.

\subsection{Velocity Dispersion Measurements}

To measure the velocity dispersion for each galaxy we followed the method described in \citet{davies93}, using a series of G- and K-type star spectra obtained with the same instrumental configuration as the galaxy spectra.

The spectrum of a galaxy is broadened due to: 1) the intrinsic velocity dispersion of the galaxy stars, $\sigma$, and 2) the instrumental resolution, $\sigma_{\rm instr}$. Thus, the final observed velocity dispersion, $\sigma_{\rm obs}$, is given by:

\begin{equation}
\sigma_{\rm obs}^2=\sigma^2+\sigma_{\rm instr}^2
\end{equation}

On the other hand, a stellar spectrum is affected only by the instrumental resolution, since a single object has zero intrinsic velocity dispersion. In this case, $\sigma=0$ and $\sigma_{\rm obs}=\sigma_{\rm instr}$. This method of measuring the intrinsic velocity dispersion of a galaxy entails broadening the stellar spectrum using a series of Gaussian filters with increasing standard deviation, $\sigma$.

Following this procedure, we estimated the intrinsic velocity dispersion for the full set of galaxies. Table \ref{spresults} presents the measured $\sigma$ values. Note that our sample includes massive and very massive objects since we are covering the range 220~km~s$^{-1}<\sigma<450$~km~s$^{-1}$.

\subsection{Ages and Metallicities}

To derive mean luminosity-weighted ages and metallicities, we compared selected absorption line strengths with those predicted by the model of Vazdekis et al. (2006, in preparation). This model provides flux-calibrated spectra in the wavelength range $\lambda\lambda~3500-7500$ \AA, at a resolution of 2~\AA\ (FWHM) for single-burst stellar populations. This model is an extension of \citet{vazdekis99}, employing the new empirical stellar spectral library of \citet{sanchezblazquez06miles}. This way, we can transform synthetic spectra to the resolution and dispersion of the galaxy spectra instead of the opposite, as required when working in the Lick/IDS system.

\begin{figure*}
\resizebox{\hsize}{!}{\includegraphics{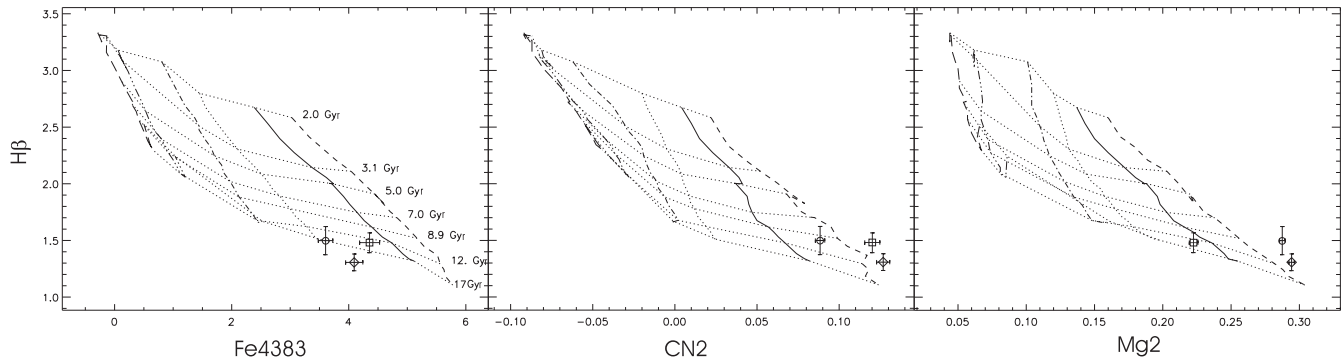}}
\caption{Examples of the method used to derive ages and metallicities. The plot shows the age index H$\beta$ versus the metallicity indicators Fe4383, CN2 and Mg$_2$ for the grid of models of Vazdekis et al. (2006, in preparation). Horizontal dotted lines represent models of constant age whose ages are labelled in the plot. Vertical dashed, solid, dotted, dashed-dotted, dashed-dotted-dotted and long-dashed lines represent contours of constant [M/H]~$=+0.2, 0.0, -0.4, -0.7, -1.2$ and $-1.7$ dex, respectively. Diamonds, open circles and squares indicate the values for a representative galaxy of A115, A655 and A963, respectively, with a common $\sigma=310$~km~s$^{-1}$. Note that all synthetic model spectra were broadened to the instrumental resolution of the observed spectra and the velocity dispersion of these particular galaxies ($\sigma_{\rm obs}=350$~km~s$^{-1}$).}
\label{modelgrid}
\end{figure*}

Once the model spectra are transformed to yield the intrumental conditions of resolution and dispersion of the observed spectrum, we measure pairs of indices in both sets of data (observed and synthetic). These pairs are formed by an index which is sensitive {\em almost} only to variations in age, and other which is sensitive {\em almost} only to variations in the abundance of a given element. Then, we obtain nearly orthogonal grids of age versus metallicity which can be compared directly with the values of the same pair of indices measured in the observed spectrum. Figure \ref{modelgrid} illustrates this method. The vertical axis shows the age indicator H$\beta$ while the three horizontal axes show the metallicity indicators Fe4383, CN2 and Mg$_2$, respectively. By comparing the model grids with the measurements in the observed galaxy spectrum we can estimate metallicities and mean luminosity-weighted ages. Note that the full set of synthetic model spectra is broadened to the velocity dispersion value of each galaxy, i.e. we create individual grids suitable for each galaxy. This way we do not need to correct for the velocity dispersion effects, to match a fixed resolution as is the case of the commonly used Lick/IDS analysis.

The selected metallicity indices were CN2, Mg$_2$ and Fe4383 \citep[c. f.][]{wortheyetal94}. We used H$\beta$ as the age index in all cases. We chose these features because of their low sensitivity to variations in signal-to-noise \citep{cardieletal03} and velocity dispersion (we estimated  $\Delta(index)/index < 0.15$, for $\Delta(\sigma)$ = 300~km~s$^{-1}$). This way we minimize possible radial variations in the index value, as $\sigma$ may vary as a function of the galactocentric radius \citep[see][]{carretero04}.

We will refer to the metallicities derived in the diagrams CN2--H$\beta$, Mg$_2$--H$\beta$ and Fe4383--H$\beta$ as [CN/H], [Mg/H] and [Fe/H], respectively. Since the CN2 index is strongly dominated by C and N, the Mg$_2$ index is governed by Mg and the Fe4383 index is driven by Fe \citep{tripiccobell95}, these metallicities must be close to the real relative abundance of each element. Given the values of the former relative abundances, we can obtain estimates of the abundance ratios [CN/Fe] and [Mg/Fe] for each galaxy.

The bulk of our galaxy sample lies with in the bottom-right corner of the model grids shown on Figure \ref{modelgrid}. However, note that an extrapolation of the model grids is required for some galaxies to obtain the abundances of CN and Mg, since the models extend only to [M/H] = 0.2. Also, it is worth noting that the data points for certain galaxies fall below the model grids, which can be attributed to the fact that absolute age determination is subject to model zeropoints \citep[see][]{vazdekisetal01b,schiavonetal02}. We neglect the possible effect of nebular emission on H$_\beta$ since no [O~III]~$\lambda$5007 emission is detected in our galaxy spectra. Nevertheless, assuming an upper H$_\beta$ emission correction of $\sim0.5$~\AA\ \citep{kuntschneretal02}, although this would give rise to a significant reduction in the mean age of the stellar populations of the oldest galaxies, the net effect on the abundance ratios would be no more than $\sim0.05$ dex, for the most affected galaxies in a cluster. This correction for the H$\beta$ index is larger than that obtained by varying the model prescriptions, e.g. with $\alpha$-enhanced isochrones and atomic diffusion included \citep{vazdekisetal01b}, used to derive ages in better agreement with the current age of the Universe, for the oldest galaxies in our sample. 

Finally, Table \ref{spresults} lists the estimates of the values of the relative abundances [CN/H], [Mg/H] and [Fe/H] and the mean luminosity-weighted ages for each galaxy in our sample.

\begin{table*}
\begin{minipage}{155mm}
\caption{Determined values of the velocity dispersion, mean luminosity-weighted age and relative abundances for our galaxy sample.}
\label{spresults}
\begin{tabular}{cccccccccc}
\hline
Galaxy ID & $\sigma$ (km~s$^{-1}$) & Log(age) & $\Delta$Log(age) & [CN/H] & $\Delta$[CN/H] & [Mg/H] & $\Delta$[Mg/H] & [Fe/H] & $\Delta$[Fe/H]\\
\hline
A$115-1$    &  450 &  10.09 & 0.35 &  0.72 & 0.11 &  0.49 & 0.17 &  0.27 & 0.16 \\
A$115-2$    &  265 &   9.90 & 0.28 &  0.57 & 0.11 &  0.72 & 0.26 &  0.05 & 0.17 \\
A$115-3$    &  355 &  10.17 & 0.26 &  0.76 & 0.08 &  0.66 & 0.13 &  0.18 & 0.11 \\
A$115-4$    &  310 &  10.16 & 0.29 &  0.34 & 0.11 &  0.34 & 0.13 & -0.21 & 0.14 \\
A$115-5$    &  370 &  10.01 & 0.31 &  1.23 & 0.26 &  0.63 & 0.28 &  0.08 & 0.18 \\
A$115-6$    &  330 &  10.18 & 0.22 &  0.04 & 0.12 & -0.09 & 0.12 &  0.03 & 0.12 \\
A$655-1$    &  315 &  10.05 & 0.36 &  0.19 & 0.29 &  0.70 & 0.22 &  0.06 & 0.33 \\
A$655-2$    &  310 &  10.11 & 0.31 &  0.17 & 0.22 &  0.58 & 0.30 & -0.57 & 0.19 \\
A$655-3$    &  250 &  10.24 & 0.00 &  0.45 & 0.14 &  0.35 & 0.10 &  0.25 & 0.17 \\
A$655-4$    &  435 &  10.12 & 0.32 &  0.62 & 0.10 &  0.47 & 0.08 &  0.05 & 0.13 \\
A$655-5$    &  295 &  10.10 & 0.31 & -0.27 & 0.15 &  0.79 & 0.30 &  0.24 & 0.22 \\
A$655-6$    &  320 &  10.11 & 0.34 & -0.49 & 0.15 &  0.03 & 0.10 &  0.27 & 0.12 \\
A$655-7$    &  260 &  10.13 & 0.31 & -0.25 & 0.14 &  0.34 & 0.09 & -0.17 & 0.14 \\
A$655-8$    &  335 &  10.13 & 0.29 &  0.31 & 0.12 &  0.20 & 0.12 & -0.34 & 0.11 \\
A$655-9$    &  325 &  10.09 & 0.32 &  0.06 & 0.11 &  0.79 & 0.20 & -0.31 & 0.12 \\
A$655-10$   &  225 &  10.10 & 0.35 & -0.51 & 0.14 &  0.00 & 0.09 & -0.69 & 0.15 \\
A$963-1$    &  275 &  10.00 & 0.39 & -0.18 & 0.14 & -0.18 & 0.08 & -0.05 & 0.08 \\
A$963-2$    &  255 &  10.15 & 0.30 &  0.18 & 0.22 &  0.31 & 0.49 & -0.58 & 0.20 \\
A$963-3$    &  425 &   9.70 & 0.28 &  0.81 & 0.10 &  0.87 & 0.13 & -0.09 & 0.16 \\
A$963-4$    &  410 &  10.17 & 0.23 &  1.11 & 0.18 &  0.68 & 0.18 &  0.49 & 0.28 \\
A$963-5$    &  310 &  10.20 & 0.21 &  0.47 & 0.14 & -0.14 & 0.11 & -0.15 & 0.12 \\
A$963-6$    &  220 &  10.12 & 0.35 &  0.85 & 0.10 &  0.97 & 0.09 &  0.46 & 0.25 \\
A$2111-1$   &  315 &  10.14 & 0.29 &  0.68 & 0.14 &  1.30 & 0.17 & -0.06 & 0.14 \\
A$2111-2$   &  420 &  10.20 & 0.23 &  0.94 & 0.08 &  0.65 & 0.08 & -0.18 & 0.10 \\
A$2111-3$   &  220 &  10.05 & 0.38 &  0.94 & 0.11 &  0.64 & 0.10 &  0.07 & 0.17 \\
A$2111-4$   &  385 &  10.19 & 0.24 &  1.30 & 0.08 &  1.30 & 0.08 &  0.04 & 0.12 \\
A$2111-5$   &  230 &   9.85 & 0.38 & -0.36 & 0.54 &  0.35 & 0.13 &  0.11 & 0.15 \\
\hline
\end{tabular}
\end{minipage}
\end{table*}

\subsection{Results}

\subsubsection{Correlations with the Galaxy Mass}

\begin{figure*}
\includegraphics[width=110mm]{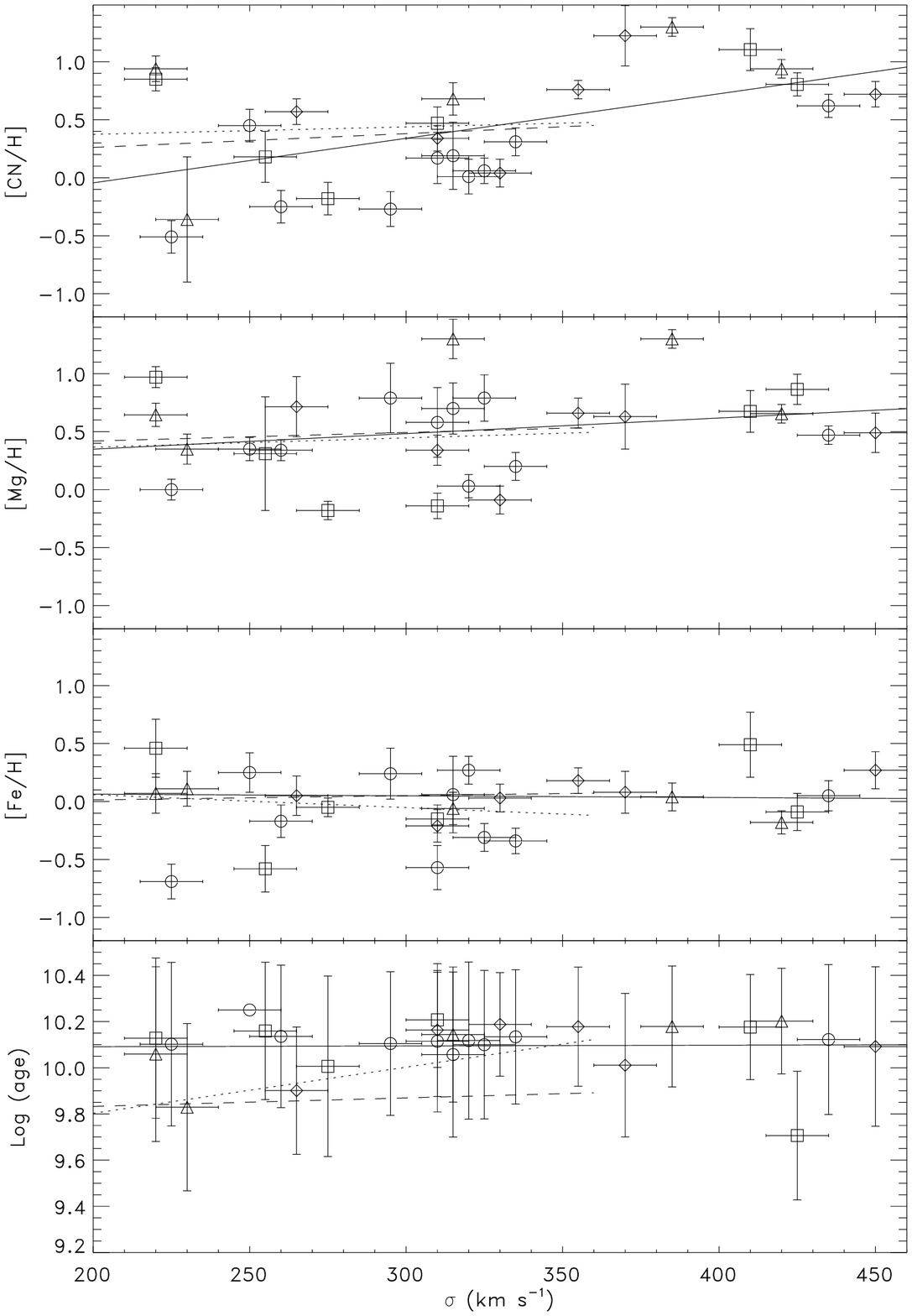}
\vspace{1cm}
\caption{Relations between the relative abundances [CN/H], [Mg/H], [Fe/H], and mean luminosity-weighted ages and the velocity dispersion of the galaxies, $\sigma$. Diamonds: A115 galaxies. Open circles: A655 galaxies. Open squares: A963 galaxies. Open triangles: A2111 galaxies. The solid line corresponds to the error-weighted linear fitting to our data. Dotted line and dashed line corresponds to the relations obtained for Virgo cluster galaxies and Coma cluster galaxies, respectively \citep[taken from][]{sanchezblazquezetal06}. It is noteworthy that the slope of the relations increases with the environment density for [CN/H]$-\sigma$, while it is almost constant in the case of [Mg/H]$-\sigma$ and [Fe/H]$-\sigma$.}
\label{abundancessigma}
\end{figure*}

Once we have obtained estimates of the relative abundances of selected elements, and the mean luminosity-weighted age for each galaxy, we studied their relation to the galactic mass, via the velocity dispersion. Figure \ref{abundancessigma} presents the relations [CN/H]$-\sigma$, [Mg/H]$-\sigma$, [Fe/H]$-\sigma$ and Log(age)$-\sigma$ for our full sample. The solid line in each panel corresponds to an error-weighted linear fit to the data. We can see that the abundances [CN/H] and [Mg/H] increase with $\sigma$. On the contrary, we find no clear trends with $\sigma$ for [Fe/H] or the age.

In order to compare our results from galaxies in very rich clusters with those from galaxies within less dense environments, we have also plotted on Figure \ref{abundancessigma} the results obtained by \citet{sanchezblazquezetal06}, who analyzed a sample of 98 early-type galaxies belonging to the Virgo and Coma clusters. The dotted line in each panel corresponds to the relation found for the Virgo cluster galaxies, while the dashed line corresponds to the Coma cluster galaxies. Note that the sigma coverage is limited to $\sigma<350$~km~s$^{-1}$ for the Virgo and Coma cluster samples.

In agreement with our findings, [CN/H] and [Mg/H] are correlated with $\sigma$ for the Virgo and Coma cluster galaxies. Also, no clear trend is found for [Fe/H]$-\sigma$. When we consider the age, the result differs from Virgo to Coma: while Virgo galaxies show a clear trend with $\sigma$, none is found for the denser Coma cluster. In particular, we observe that the behaviour of the Log(age)$-\sigma$ relation is very similar for the Coma galaxies and for our sample.

Note that, in the bottom panel of Figure \ref{abundancessigma}, there is an offset between the Log(age)$-\sigma$ relation for our cluster sample and the Virgo and Coma ones. This is not a critical issue since we are only interested in the slopes of the correlations, and not in their absolute values. Nevertheless, the analysis of the relative age values does give useful information. This is also the case for derived metallicity values: we do not expect to obtain absolute values of [M/H], since we based our analysis on the relative ones.

\begin{figure*}
\includegraphics[width=110mm]{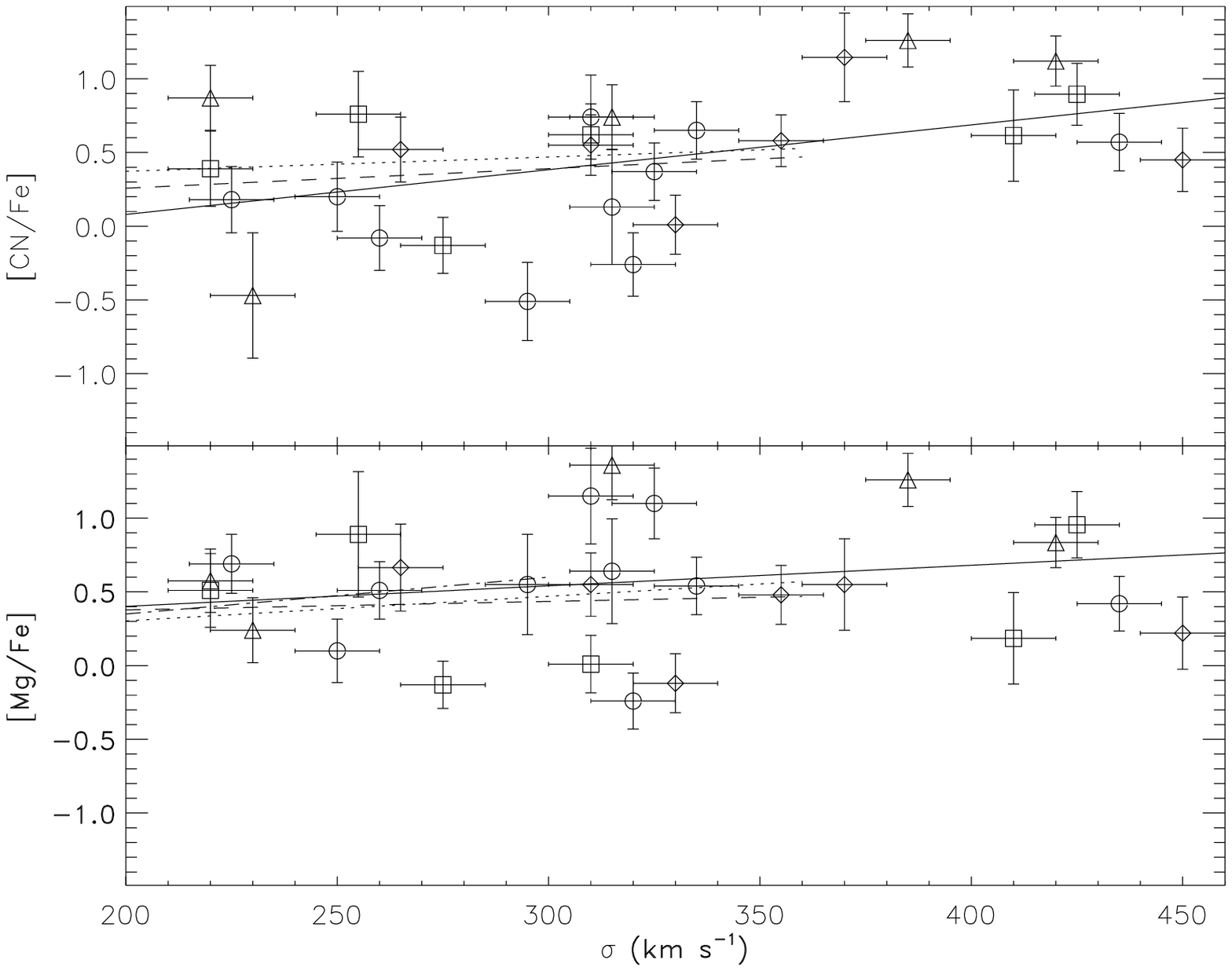}
\vspace{1cm}
\caption{Abundance ratios [CN/Fe] and [Mg/Fe] versus the velocity dispersion of the galaxies. Diamonds: A115 galaxies. Open circles: A655 galaxies. Open squares: A963 galaxies. Open triangles: A2111 galaxies. The solid line corresponds to the error-weighted linear fitting to our data. Dotted line and dashed line corresponds to the relations obtained for Virgo cluster galaxies and Coma cluster galaxies, respectively \citep[taken from][]{sanchezblazquezetal06}. Dashed-dotted line in bottom panel corresponds to the [Mg/Fe]$-\sigma$ relation for Virgo cluster galaxies, by \citet{vazdekistrujillo04}. Note that the slope of the [CN/Fe]$-\sigma$ relation increases with the density of the environment, but the [Mg/Fe]$-\sigma$ remains almost constant.}
\label{abundanceratios}
\end{figure*}

We have focused so far on the relative abundances of selected elements. However, in order to obtain information concerning characteristic formation timescales for the bulk of their stellar populations, we computed the abundance ratios [CN/Fe] and [Mg/Fe]. Since each element was ejected into the ISM by different types of stars at different epochs, their relative abundance values will be very useful for constraining the star formation history. Figure \ref{abundanceratios} presents the relations of the abundance ratios [CN/Fe] and [Mg/Fe] with the velocity dispersion in our sample. The solid line in each panel corresponds to an error-weighted linear fit to the data. We observe positive trends with $\sigma$ for both abundance ratios [CN/Fe] and [Mg/Fe].

As in the previous relations, we compared our results with those obtained in Virgo and Coma clusters. Figure \ref{abundanceratios} shows the results. The dotted line in each panel corresponds to the relation found for Virgo cluster galaxies, while the dashed line corresponds to Coma cluster galaxies. In this case, we also see correlations of both [CN/Fe] and [Mg/Fe] with $\sigma$. We have also compared our result with the [Mg/Fe]$-\sigma$ relation for Virgo cluster galaxies found by \citet{vazdekistrujillo04}, which is represented in bottom panel of Figure \ref{abundanceratios} by a dashed-dotted line.

\subsubsection{Correlations with the Environment}

\begin{figure*}
\includegraphics[width=140mm]{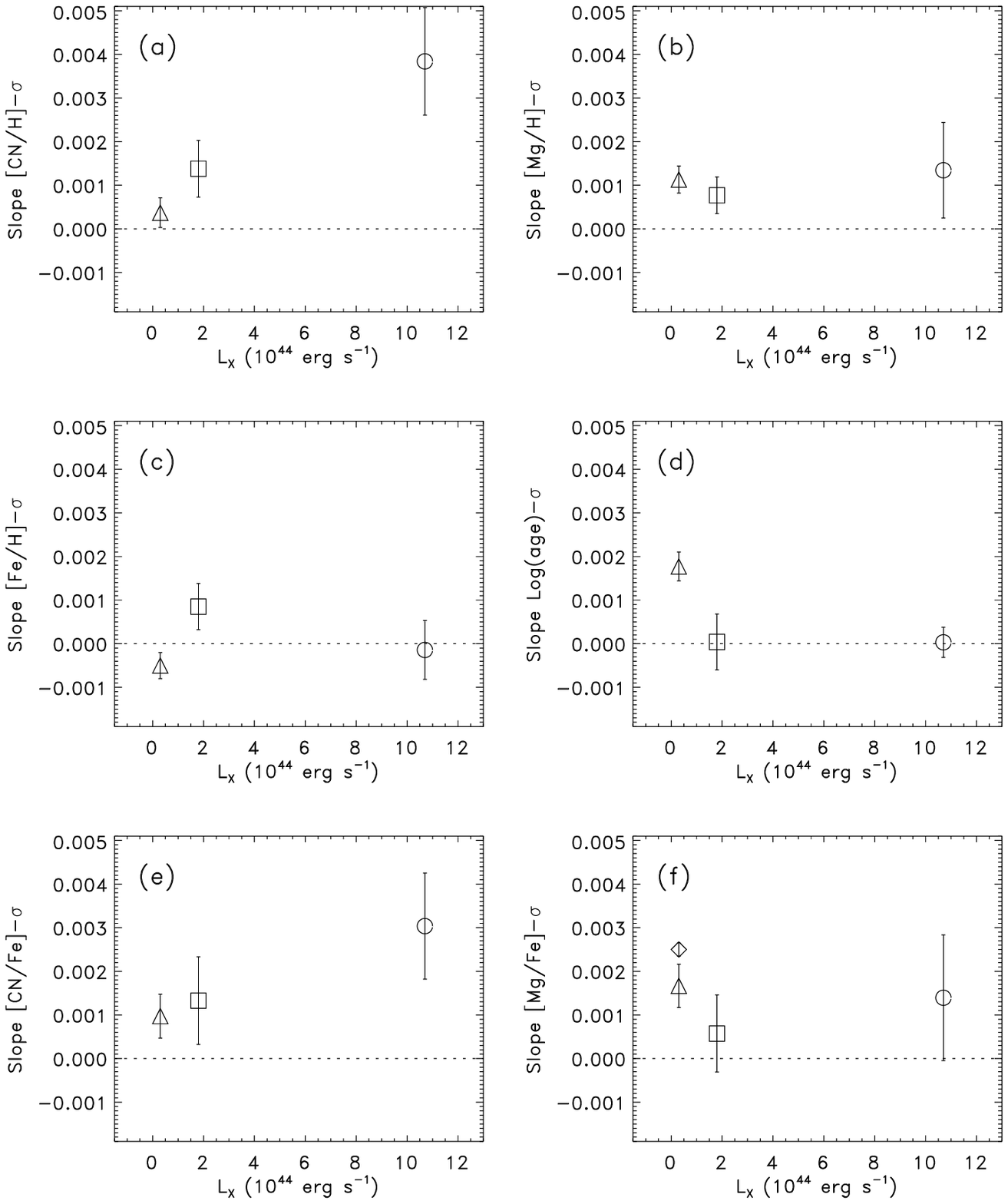}
\caption{Slopes of the linear fits for the relations [CN/H]$-\sigma$ (a), [Mg/H]$-\sigma$ (b), [Fe/H]$-\sigma$ (c), Log(age)$-\sigma$ (d), [CN/Fe]$-\sigma$ (e), [Mg/Fe]$-\sigma$ (f), versus the X-ray luminosity of the host clusters. Horizontal dotted line indicates a zero slope, i.e., the case in which metallicity and $\sigma$ are not correlated. Triangles and squares correspond to Virgo and Coma cluster slopes, respectively \citep[both taken from][]{sanchezblazquezetal06}. Open circles correspond to our results. We considered the average $L_{\rm X}$ within our cluster sample as the representative value. The diamond in panel (f) indicates the value for the Virgo cluster [Mg/Fe]$-\sigma$ slope obtained by \citet{vazdekistrujillo04}. There exists a trend for the CN abundance to increase with the cluster mass, as well as the [CN/Fe] abundance ratio. No relation with the environment is found for the other elements.}
\label{slopes}
\end{figure*}

Figure \ref{slopes} presents the slopes of the above Z$-\sigma$ relations (where Z represents any of the abundance ratios above) as a function of the cluster X-ray luminosity, which is a good indicator of the cluster mass \citep[e.g.][]{ettorietal02,donahueetal03,shimizuetal03} and a useful representative of its richness class. This way we can compare the strength of the full set of relations as a function of the environment. Figures \ref{slopes}a to \ref{slopes}d present the slopes of the relations [CN/H]$-\sigma$, [Mg/H]$-\sigma$, [Fe/H]$-\sigma$ and Log(age)$-\sigma$ as functions of $L_{\rm X}$, for our sample and for the Virgo and Coma cluster galaxies. For the slope of the [CN/H]$-\sigma$ relation, we observe a clear trend of the slopes to increase with the density of the environment. On the contrary, when considering the slopes of the [Mg/H]$-\sigma$ relation we do not find a clear trend with the environment, all slopes being almost constant and positive. In addition, the [Fe/H]$-\sigma$ relation presents slopes close to zero independently of the properties of the host cluster. Different behaviour is found for the relation of the slope of Log(age)$-\sigma$ as a function of the environment. In this case, our results suggest that this slope is anticorrelated with the X-ray luminosity.

Finally, Figures \ref{slopes}e and \ref{slopes}f present the dependence of the slopes of the [CN/Fe]$-\sigma$ and [Mg/Fe]$-\sigma$ relations, on the cluster $L_{\rm X}$. We can observe that, for both abundance ratios, the slopes are positive. In addition, the slope of the [CN/Fe]$-\sigma$ relation increases with the cluster mass, while the slope of the [Mg/Fe]$-\sigma$ relation is nearly constant, though positive, within the errors.

\section{Discussion}

In this Section we will try to explain within a common scenario all the trends found. In particular, the correlations can be interpreted in terms of the different formation timescales for each chemical element, and the different star formation histories of elliptical galaxies as a function of their environment.

Magnesium is ejected into the interstellar medium (ISM) by Type II supernovae (SNe II) on short timescales \citep[$<10$~Myr;][]{faberetal92,matteucci94}. On the other hand, the iron-peak elements are the products of SNe~Ia, which occur on timescales of $\sim1$~Gyr. Between the two extremes, although there are recent suggestions that most of the C comes from massive stars \citep{akermanetal04}, C and N are mainly ejected into the ISM by low- and intermediate-mass stars \citep{renzinivoli81,chiappinietal03}, leading to CN formation on timescales longer than that for Mg but shorter than that for Fe. Several authors \citep[e.g.][]{ellisetal97,stanfordetal98} argue that early-type galaxies are old and passively evolving systems. The luminosity-weighted ages derived from our model grids confirm that the galaxies are significantly older than the formation timescales of the different species. So, if we find substantial differences in the abundance ratios of these elements which depend on the physical properties of the environment, these must be due to the fact that galaxies are assembled on different timescales as a function of their environment.

We will focus our discussion on the following three aspects:
\begin{itemize}
\item[1)] The comparison of the relative abundances and abundance ratios measured in galaxies of the {\em same} mass (i.e. equal velocity disperion) in {\em different} environments.
\item[2)] The comparison of the stellar population properties measured in galaxies of {\em different} mass within the {\em same} environment.
\item[3)] The dependence of above relations as functions of the galaxy mass, with the environment properties.
\end{itemize}

First, when inspecting Figures \ref{abundancessigma} and \ref{abundanceratios} we observe that, for a given $\sigma$, the relative abundance [CN/H] increases with the cluster X-ray luminosity, whereas [Mg/H] and [Fe/H] do not. Something similar happens with the abundance ratios: while [CN/Fe] increases as a function of the environment density, the [Mg/Fe] ratio keeps constant and {\em positive}. We interprete the fact that [Mg/Fe] is positive and constant in terms of the formation timescales of the stars in ellipticals, as a function of the environment.

If we consider the formation timescales for Mg and Fe stated above we can estimate that, whatever the environment is, all stellar populations in ellipticals should be already formed in less than $\sim1$~Gyr, since this is the time needed by SNe~Ia to pollute the ISM with iron-peak elements. But, if we consider species with less separated formation timescales, such as Fe and CN, we do find differences in the ratio [CN/Fe] as a function of the environment, for a given $\sigma$. That suggests a different formation timescale for the stars in ellipticals related to the environment properties, in the sense that the more dense the environment is the shorter the formation timescales of the stars are (with an upper limit of $\sim1$~Gyr). This is in agreeement with a number of recent results \citep[e.g.][]{carretero04,thomas05,bernardi06}.

Second, for our sample of galaxies in very rich clusters, we do find positive trends between [CN/H], [Mg/H], [CN/Fe], [Mg/Fe] and $\sigma$. On the contrary, no correlations are found between [Fe/H] or Log(age) with the velocity dispersion. These relations are explained within a framework where more massive galaxies produce more metals \citep{kodama97,chiosi02,matteucci03}. Our results are compatible with those of \citet{trageretal00}, who found a relation between the abundance of some $\alpha$-elements, including Mg, and the velocity dispersion for a sample of old cluster galaxies. Moreover, \citet{vazdekisetal01b} and \citet{yamada06} suggested that the CMR is driven essentially by metallicity and, in particular, by the abundance of the $\alpha$-enhanced elements.

The Log(age)$-\sigma$ relation presented in Figure \ref{abundancessigma} and the slopes of Figure \ref{slopes}d show that  the mean stellar age is nearly constant for all galaxies within rich clusters. This is not the case for the Virgo cluster galaxies, which present bigger ages for the most massive objects. This is due to the fact that smaller galaxies in low density environments present a larger scatter in the age$-\sigma$ relation \citep[see][]{sanchezblazquezetal06}. In fact, the slope in age$-\sigma$ for the Virgo cluster is also near zero if the sample is limited to galaxies with sigma larger than 200~km~s$^{-1}$.

Third, it is noteworthy that the slope of the relation [CN/Fe]$-\sigma$ increases with the cluster X-ray luminosity while the slope of the [Mg/Fe]$-\sigma$ is constant with the cluster mass (see Figure \ref{slopes}). Because of that, we can only find environment-related [CN/Fe] differences in those galaxies with $\sigma<300$~km~s$^{-1}$ whereas, for more massive galaxies, the [CN/Fe] values are almost equal independently of the cluster properties. This means that the ratio [CN/Fe] in most massive galaxies is less sensitive to the environment than in intermediate- and low-mass galaxies, suggesting that star formation histories are less environment dependent as the galaxy mass increases. Therefore, [CN/Fe] appears to be an appropriate ``chemical clock'' for ellipticals with $\sigma<300$~km~s$^{-1}$ but, when we move to very massive galaxies, the formation timescales differ so gently from one environment to another that this ``clock'' turns out to be insufficiently sensitive. In order to disentangle the different formation timescales between very massive galaxies in different environments, it is neccesary to find out a more accurate indicator.

Our results suggest an overall picture in which massive elliptical galaxies in very rich clusters are old systems, with very short star formation histories, and which have been passively evolving since, at least, $z\sim2$. This is in agreement with a number recent results \citep[e.g.][]{labee05,kriek06,vandokkum06}. In comparison with less dense environments, the stars in elliptical galaxies of rich clusters must have formed at slightly earlier epochs and on a slightly shorter timescales.

\section{Conclusions}

We have performed a detailed stellar population analysis of a sample of 27 massive elliptical galaxies  (220~km~s$^{-1}<\sigma<450$~km~s$^{-1}$) within 4 very rich Abell clusters: A115, A655, A963 and A2111. From the absorption indices H$\beta$, CN2, Mg$_2$ and Fe4383 we derived mean luminosity-weighted ages, relative abundances [CN/H], [Mg/H] and [Fe/H], and the abundance ratios [CN/Fe] and [Mg/Fe]. To obtain these results we used our new, high-resolution stellar population models (Vazdekis et al 2006, in preparation). We obtained the dependences of all parameters above on the galaxy velocity dispersion.

Given the different formation timescales for each element, we have shown that the abundance ratios [CN/Fe] and [Mg/Fe] are the key ``chemical clocks'' for infering the star formation history timescales in ellipticals. In particular, [Mg/Fe] provides an upper limit for those formation timescales, while [CN/Fe] apperars to be the most suitable ``clock'' to discriminate finely among them. This is true for those galaxies with $\sigma<300$~km~s$^{-1}$ as, for more massive galaxies, the [CN/Fe] ratio loses its ''timing'' sensitivity.

Our [Mg/Fe] results are compatible with a scenario in which the stellar populations of elliptical galaxies, independently of their environment and mass, had formation timescales shorter than $\sim1$~Gyr. This result implies that massive elliptical galaxies have evolved passively since, at least, $z\sim2$.

On the other hand, the information provided by the [CN/Fe] analysis suggests that, for a given galaxy mass (i.e. a given velocity dispersion), the star formation history must be shorter in those galaxies belonging to more dense environments. Furthermore, we found that these timescales differences become smaller for the most massive cluster galaxies ($\sigma>300$~km~s$^{-1}$), so they cannot be traced by the [CN/Fe] abundance ratio.

Finally, we found that the mass-metallicity relation is steeper as the density of the environment increases.

Our results do not allow us, at this point, to decide between the two principal scenarios, hierarchical or monolithic, for galaxy formation, but do set certain strong constraints. Specifically, we have shown that galaxy assembly timescales for ellipticals are short and environment dependent, and this must be addressed by any valid theory of galaxy formation.

\section*{Acknowledgments}

The authors thank J. Cenarro for useful comments concerning the data reduction. A. V. is a Ram\'on y Cajal Fellow of the Spanish Ministry of Education and Science. Based on observations made with the Italian Telescopio Nazionale Galileo (TNG) operated on the island of La Palma by the Fundaci\'on Galileo Galilei of the INAF (Istituto Nazionale di Astrofisica) at the Spanish Observatorio del Roque de los Muchachos of the Instituto de Astrofisica de Canarias. This work has been supported by the Spanish Ministry of Education and Science grants AYA2004-03059 and AYA2004-08251-C02-01.


\label{lastpage}

\end{document}